# Comparative Studies on Decentralized Multiloop PID Controller Design Using Evolutionary Algorithms

Sayan Saha, Saptarshi Das, Anindya Pakhira, Sumit Mukherjee, and Indranil Pan

*Abstract*—Decentralized PID controllers have been designed in this paper for simultaneous tracking of individual process variables in multivariable systems under step reference input. The controller design framework takes into account the minimization of a weighted sum of Integral of Time multiplied Squared Error (ITSE) and Integral of Squared Controller Output (ISCO) so as to balance the overall tracking errors for the process variables and required variation in the corresponding manipulated variables. Decentralized PID gains are tuned using three popular Evolutionary Algorithms (EAs) viz. Genetic Algorithm (GA), Evolutionary Strategy (ES) and Cultural Algorithm (CA). Credible simulation comparisons have been reported for four benchmark 2×2 multivariable processes.

*Index Terms*-Cultural Algorithm; Evolutionary Strategy; Genetic Algorithm; multivariable process control; PID controller

## I. INTRODUCTION

IN most of the industrial process control systems, generally few manipulated variables are adjusted to control a number of controlled variables. Such control systems are known as multi-variable processes and more commonly termed as multiple-input multiple-output (MIMO) systems [1]. In multivariable process control, unlike single-input, single-output (SISO) systems, change of any single manipulated variables affects more than one controlled variable, giving rise to the loop interaction [2]. The simplest MIMO processes which are often referred in multivariable process control literatures consist of two inputs and two outputs (TITO) [3]. Conventional method of designing PID type controllers for such MIMO processes require correct pairing of one manipulated variable to one controlled variable, to avoid poor controller performance and reduced stability margins, which can be achieved by means of Relative Gain Array (RGA) approach. Other improved measures of loop interaction, necessary and sufficient conditions for pairing, control structure selection etc. have been thoroughly reviewed in [2], though the RGA based loop interaction analysis still dominates the process control industries. Only steady-state information about the process gain matrix is required to develop a RGA, which provides a measure of process interactions between the manipulated variables and controlled variables. The most effective pairing can be achieved if a manipulated variable is used to monitor a controlled variable with which its measure of interaction is highest, preferably, close to unity. This allows pairing of a single controlled variable with a single manipulated variable via a feedback controller for two such loops of TITO process as in Fig. 1.

Thus in decentralized PID control of 2×2 MIMO systems, the control system consists of two such controllers. Each of them takes care of a single loop only and the interaction between the two loops is greatly reduced unlike the centralized PID control where similar 2×2 controller structure is assumed to stabilize a TITO process [4]-[5]. However, the decentralized controller design can be easily applied if the loops do not heavily interact with each other i.e. the corresponding RGA should have dominating principal diagonal. If the loop interaction changes the process gains of the individual loops considerably, then well-tuned controller for the individual loops fail to keep the controlled variables at their respective set-points. In such cases, pairing of any manipulated variable with any controlled variable results in poor controller performance. One method to overcome this problem, as attempted in this paper, is to tune both the loops simultaneously instead of tuning decoupled loops individually. This is because in the later case when one loop is being tuned, the controlled variable of the other loop is not traced. Thus it is likely to deviate from the set-point in case of large loop interaction.

One of the conventional methods of tuning SISO control loops in frequency domain is user-specified gain and phase-margin (GPM) methods. However, in case of MIMO systems, the number of control loops is two or more, resulting in a number of pairs of gain-margins and phase-margins i.e. one pair for each loop, calculated by opening that individual loop only. Thus, the presence of multiple loops makes it impossible to correctly assign a particular gain-margin and phase-margin to a MIMO system and hence tuning of MIMO systems in frequency domain becomes very difficult. Huang *et al.* [6] have proposed a methodology to decompose MIMO processes as several effective open loop processes and applied the GPM method for those individual loops. This method may be used to monitor the speed of response for each loop of the MIMO systems. However this does not take into consideration the required variation in the manipulated variables or control signal which is required in order to limit the actuator size and variation of the manipulated variables. The time domain integral performance index based PID controller tuning for

S. Saha and A. Pakhira are with the Department of Instrumentation and Electronics Engineering, Jadavpur University, Salt-Lake Campus, LB-8, Sector 3, Kolkata-700098, India.
S. Das, S. Mukherjee, and I. Pan are with the Department of Power Engineering, Jadavpur University, Salt-Lake Campus, LB-8, Sector 3, Kolkata-700098, India. (e-mail: saptarshi@pe.jusl.ac.in).

MIMO processes, as attempted in this paper was first introduced by Zhuang and Atherton [3] whereas the present paper improves this technique by taking both loop error index (ITSE) and required controller effort (ISCO) simultaneously into consideration. In case of optimal tuning of PID controllers to handle MIMO systems, if each of the controlled variables is enforced to track a set-point using some optimization technique, while ignoring the variation in the other controlled variables, it becomes almost impossible to track both the controlled variables to desired set-points with the tuned controller parameters. In this paper, multivariable PID controllers are designed by simultaneous optimal tuning of all the control loops for 2×2 multivariable benchmark processes with high loop interaction viz. Wood and Berry (WB), Vinante and Luyben (VL), Wardle and Wood (WW), and Ogunnaike and Ray (OR) [6]. Optimum tuning is attempted using three well known evolutionary algorithms viz. Genetic Algorithm (GA), Evolutionary Strategy (ES) and Cultural Algorithm (CA) for simultaneous tracking of all the controlled variables around desired set-points, implemented as unit step reference inputs.

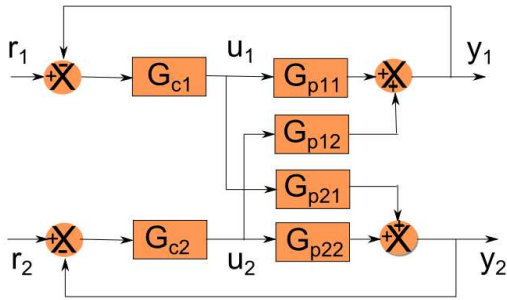

Figure 1. Schematic of the decentralized PID control structure for benchmark TITO processes.

Similar computational intelligent and optimization based attempts have been made in few contemporary literatures for the tuning of multi-loop PID controllers e.g. Chang *et al.* [7] proposed an on-line scheme for PID controller design for multivariable processes, using auto-tuning neurons employing hyperbolic tangent activation function. Iruthayarajan and Baskar [8] and Chang [9] used evolutionary algorithms and multi-crossover genetic algorithm to minimize the summed integrated absolute error (IAE) for each loop while tuning the PID controller parameters. Rajabioun *et al.* [10] designed a decentralized PID controller by minimizing total IAE for all loops using colonial competitive algorithm. Han *et al.* [11] tuned PID controller based on a closed loop particle swarm optimizer (PSO) algorithm. Zhao *et al.* [12] minimized integral square error (ISE) employing two-*lbests* based PSO for designing robust PID controller for MIMO systems. This paper puts forward a new methodology of tuning MIMO control loops taking set-point tracking and controller effort both into consideration and comparison is made between three different EA based decentralized PID controller tuning.

Rest of the paper is organized as follows. Section II describes the basics of time domain tuning of PID controllers to handle MIMO processes. Brief description of three EAs, used for controller tuning is presented in Section III. Comparison of control performances for the benchmark TITO processes are outlined in Section IV. The paper ends with the conclusion as Section V, followed by the references.

## II. TIME-DOMAIN OPTIMUM PID CONTROLLER DESIGN FOR MIMO PROCESSES

### A. Proposed Approach of PID Controller Tuning for Multivariable Processes

The controller for each loop has been considered as the PID type in parallel structure since PID controllers still dominates process industries due to their simplicity, robustness and ease of implementation. The PID controller parameters $\{K_p, K_i, K_d\}$ for each loop of the multi-variable process is tuned in an optimal fashion so as to keep each of the controlled variables at their set-points, irrespective of any change in set-points of other controlled variables. In many process control applications, large variations of manipulated variables are not allowed to keep the physical strain of the final control element within limits. The control action is hence limited to minimize the deviations of manipulated variables. Evolutionary algorithms are used to minimize the objective function which takes into account both the deviation of the controller output and that of the controlled variable, given by:

$$J = \int_0^\infty \left[ w_1 \cdot t \cdot e^2(t) + w_2 \cdot u^2(t) \right] dt \qquad (1)$$

The first term in the above expression corresponds to the ITSE which minimizes the overshoot and settling time, whereas the second term denotes the ISCO. The two weights $\{w_1, w_2\}$ balances the impact of control loop error (oscillation and/or sluggishness) and control signal (larger actuator size and chance of integral wind-up) and both have been chosen to be unity in the present simulation study, indicating same penalty for large magnitude ITSE and ISCO. Evolutionary Algorithms (EA) have now been employed to obtain optimum PID controller parameters within the range of $\{K_p, K_i, K_d\} \in [-6, 6]$ to minimize the objective function (1). In the proposed approach the transfer function matrix of any process need not necessarily be a square one, a criterion that must be met if de-couplers are to be used for controlling the multivariable process. However, since most of the classical test-bench problems have 2×2 transfer function matrices, the simulation study has been limited to 2×2 multi-variable processes only in this paper. The same principle can be used to tune higher dimensional MIMO processes. However, in such cases the time required for such an optimization with large number of decision variables, representing the controller parameters will also increase. Since the proposed multi-loop controller tuning methodology is offline, thus computational complexity and required time for the EA to converge is not a major concern.

### B. Multi-loop Test-bench Processes

In order to test the effectiveness of the proposed tuning methodology, four different 2×2 multi-variable processes, normally encountered in process control applications, have been considered as a test-bench [6]. The transfer function matrices for Wood and Berry, Vinante and Luyben, Wardle and Wood, Ogunnaike and Ray MIMO processes are given by equations (2)-(5) respectively:

$$G_{WB}(s) = \begin{bmatrix} \dfrac{12.8e^{-s}}{16.7s+1} & \dfrac{-18.9e^{-3s}}{21s+1} \\ \dfrac{6.6e^{-7s}}{10.9s+1} & \dfrac{-19.4e^{-3s}}{14.4s+1} \end{bmatrix} \quad (2)$$

$$G_{VL}(s) = \begin{bmatrix} \dfrac{-2.2e^{-s}}{7s+1} & \dfrac{1.3e^{-0.3s}}{7s+1} \\ \dfrac{-2.8e^{-1.8s}}{9.5s+1} & \dfrac{4.3e^{-0.35s}}{9.2s+1} \end{bmatrix} \quad (3)$$

$$G_{WW}(s) = \begin{bmatrix} \dfrac{0.126e^{-6s}}{60s+1} & \dfrac{-0.101e^{-12s}}{(48s+1)(45s+1)} \\ \dfrac{0.094e^{-8s}}{38s+1} & \dfrac{-0.12e^{-8s}}{35s+1} \end{bmatrix} \quad (4)$$

$$G_{OR}(s) = \begin{bmatrix} \dfrac{22.89e^{-0.2s}}{4.572s+1} & \dfrac{-11.64e^{-0.4s}}{1.807s+1} \\ \dfrac{4.689e^{-0.2s}}{2.174s+1} & \dfrac{5.8e^{-0.4s}}{1.801s+1} \end{bmatrix} \quad (5)$$

As previously mentioned, proper pairing of manipulated variable with the controlled variable is required to minimize the effect of loop interaction as much as possible for the design of decentralized controllers for multivariable processes. The common criterion, used to obtain the knowledge of correct pairing is the relative gain array or RGA, derived from the dc gain $G_p(0)$ of the steady state process transfer matrix as:

$$RGA = G_p(0) \otimes (G_p(0)^{-1})^T \quad (6)$$

where, $\otimes$ denotes the element-by-element multiplication of the matrices. The RGA for the above four test-bench processes are:

$$RGA_{WB} = \begin{bmatrix} 2.0094 & -1.0094 \\ -1.0094 & 2.0094 \end{bmatrix} \quad (7)$$

$$RGA_{VL} = \begin{bmatrix} 1.6254 & -0.6254 \\ -0.6254 & 1.6254 \end{bmatrix} \quad (8)$$

$$RGA_{WW} = \begin{bmatrix} 2.6875 & -1.6875 \\ -1.6875 & 2.6875 \end{bmatrix} \quad (9)$$

$$RGA_{OR} = \begin{bmatrix} 0.7087 & 0.2913 \\ 0.2913 & 0.7087 \end{bmatrix} \quad (10)$$

It is observed from the RGA values of the test-bench TITO processes that the processes suffer from high loop interactions. Hence designing decentralized controllers by pairing of any manipulated variable with any controlled variable will not lead to a satisfactory performance. In this paper, we utilized simultaneous tuning of both the PID controllers at a time for considering the effect of loop interactions in the tuning phase while minimizing the objective function (1) for all loops, instead of tuning one controller as in a SISO loop.

### III. BRIEF DESCRIPTION OF THE OPTIMIZATION ALGORITHMS USED IN THE PRESENT CONTROLLER TUNING PROBLEM

In MIMO processes, it is often observed that one loop is stable while the other one is unstable when all other loops in the system except the concerned one are opened. However, the MIMO system as a whole may be stable when all the loops are closed. It is also seen that while one of the individual loops involves a positive process gain, some other loop has negative process gain. Also, some loops exhibit a leading nature while rest of the loops may have lagging characteristics. To stabilize a MIMO process, where co-existence and combination of such different dynamical behaviors are inevitable, it is important to avoid local minima while searching for optimum controller parameters. To resolve this problem evolutionary algorithms are used in this study, since they are derivative free global optimizers, capable of avoiding local minima. We have developed in-house MATLAB codes to implement Genetic Algorithm, Evolutionary Strategies and Cultural Algorithm as three representative cases of evolutionary algorithms [13].

*A. Genetic Algorithm*

Genetic Algorithm is a stochastic optimization technique based on Darwin's theory of natural selection. In GA, an initial set of probable solutions, viewed as a pool of chromosomes, is evolved over successive generations using the processes of selection, crossover and mutation, to arrive at an optimum. A fitness function is defined which measures the closeness of a chromosome to the required optimum. In each generation the 'fitness' of each individual chromosome in the population is measured and a set of most fit chromosomes is selected, based on some well-defined criteria. This is termed as selection. The selected chromosomes are then treated as parents for the reproduction of the next generation. A fraction of the parents undergo crossover, where genetic information between two chromosomes is blended. The remaining fraction undergoes random mutation. The crossover represents traversal of the search space for optimum solutions. Mutation is used to prevent pre-mature convergence and trapping into local minima/maxima. In this way, a new generation is formed, and the process is repeated. In certain GA variants, a small set of best fit individuals is directly put into the population of the next generation. This is termed elitism. Real GA, where each chromosome is represented as a vector of real values, poses as a very viable variant of GA and is specifically suited for solving optimization problems employing large continuous search spaces and has been used in the present multivariable PID controller tuning problem.

*B. Evolutionary Strategies*

Among several variants of ES, self-adapting ES is used in this study. Evolutionary strategy begins with an initial set of population consisting of real valued vectors termed as individuals, each one representing the genetic characteristics of the decision vector to be optimized. Evolution of the initial population formed by randomly generated individuals, called parents, consists of evolving of the genetic characteristics controlled by the strategy parameter ($\sigma$) which is also evolved dynamically, depending upon the performance. Strategy parameter of each individual is set to a common value of 0.5 initially. Multi-membered ES, denoted by ($\mu/\rho + \lambda$) is used for optimization. For the purpose of determining the optimized value of the decision vector, number of parents ($\mu$) and number of off-springs ($\lambda$) generated in one generation are fixed to 30. In each generation two parents ($\rho = 2$) were selected at random and recombined by a randomized real-valued crossover operator to generate an offspring. Evolution of strategy parameter is done following the modified 1/5th update rule, reported in Greenwood and Zhu [14] according to which, $\sigma$ is updated after every $n = 9$ trials depending on the number of successful mutations. Mutation of each offspring is then done by adding the zero-mean Gaussian variable with standard deviation ($\sigma$). Selection to form the new population is done by choosing the best ($\mu$) individuals out of the pool of the total ($\mu + \lambda$) individuals comprising of all the parents and offsprings depending upon their fitness value. This whole process is repeated until the required value of fitness is reached or the maximum number of generations is exceeded.

*C. Cultural Algorithm*

Cultural Algorithm has been developed by modeling how human cultures work. Culture is viewed as a vehicle for storing relevant information gathered since the start of the culture, and is available to all the subsequent generations of the society. This information can be useful for the generations to guide their problem solving activities, at the same time being dynamically modified by new information gathered by each new generation. The CA is modeled using two separate information spaces, viz. the population space and the belief space. The population space contains the set of possible solutions to the problem available in the present generation. The belief space models the actual cultural aspects. It stores information related to the problem solution that has been found till the present generation and in turn influences the evolution of the population space in subsequent generations.

Communication between the two spaces is handled by a protocol consisting of two functions: an acceptance function, which selects the set of individuals that will influence the belief space and an influence function which influences the creation of the next generation. In addition the belief space requires an update function which is basically responsible for updating the belief space when required. The belief space is composed of a few knowledge sources viz. normative knowledge, topographical knowledge, situational knowledge and history knowledge. In the present case, a variation of the CA has been used, where the evolutionary aspect is handled by a Genetic Algorithm [15]. In this Genetic Algorithm based Cultural Algorithm (GACA), the acceptance function accepts the best 25% of the population using stochastic uniform sampling, to be responsible for updating the belief space. Belief space is composed of normative and situational components. The normative knowledge component is composed of the upper and lower bounds of each of the variables among the individuals accepted. The situational knowledge is a set of the best or elite individuals. At the end of each generation the normative knowledge is updated with the bounds of the accepted individuals and the situational component is updated if necessary. Mutation in the Genetic Algorithm part is influenced by the belief space. The direction of mutation is determined by the position of current individual with respect to individual present in the situational knowledge space. Mutation is directed towards the best individual. Also, mutation range i.e. the maximum range through which the individual can be mutated, is determined by the width of the normative knowledge component for each of the variables of an individual solution. Finally, if an individual after mutation or crossover violates the normative knowledge, it is forced back into the search space dictated by the normative knowledge.

IV. SIMULATION AND RESULTS

The above discussed three EAs have now been applied to tune decentralized PID controller parameters (Fig. 1) for each of the test-bench TITO processes (2)-(5). The set-point tracking performance and required variation in manipulated variables of each loop have been shown in Fig. 2-9 respectively. Table I-II reports the best found optimum controller parameters for the two loops of each TITO process using GA, ES and CA. It can be easily inferred from figures presented that all the three algorithms viz. GA, ES and CA performs satisfactorily in simultaneous tuning of both the PID controllers for the multivariable processes with high loop interaction. From the results presented in Table III it is evident that though performance of all the three algorithms are almost same, CA outperforms the other two by some margin in all the cases.

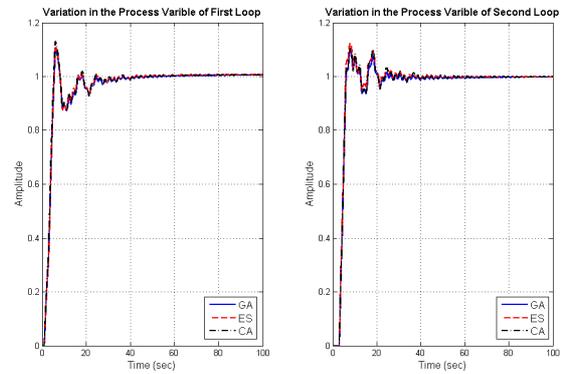

Figure 2. Variation in process variables for WB MIMO process.

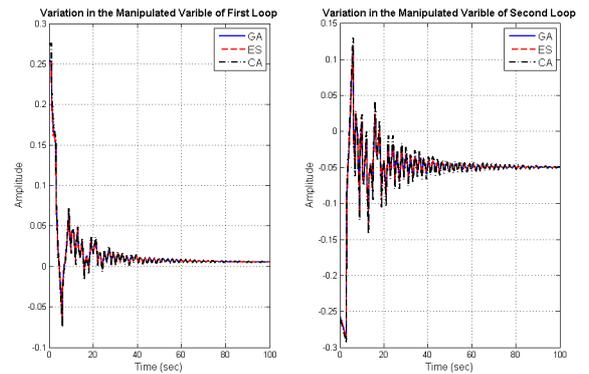

Figure 3. Variation in manipulated variables for WB MIMO process.

It is to be noted that variation in the results obtained from five independent runs of each EAs is minimum for CA. Some of the obtained controller gains are negative which is due to the fact that the MIMO process has negative transfer functions in some loops. The tuned response with the controllers, using the proposed methodology, is somewhat process dependent. From Fig. 2-3, it can be seen that the WB process has slight overshoot in tracking of process variables and oscillations in manipulated variables. From Fig. 4-5 it is observed that the

VL process gives sluggish output in one loop and slight overshoot in the other. For the WW process the settling time is very large and hence the control signal has not settled during the finite simulation time horizon (Fig. 6-7). From Figs. 8-9 it can be inferred that the OR process gives fast set-point tracking for both the loops.

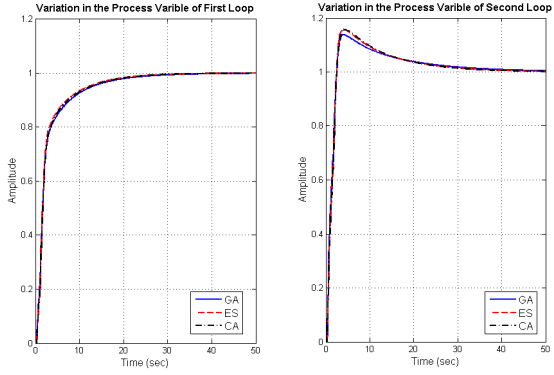

Figure 4. Variation in process variables for VL MIMO process.

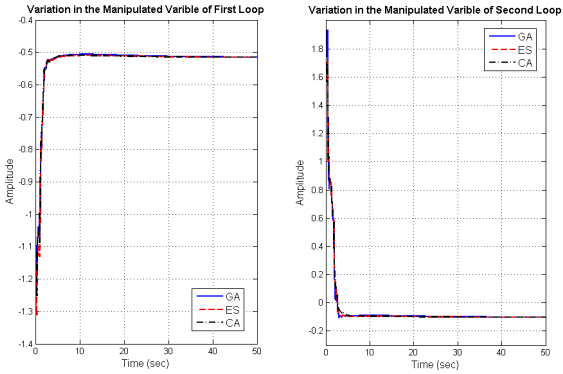

Figure 5. Variation in manipulated variables for VL MIMO process.

TABLE I. TUNED CONTROLLER PARAMETERS FOR THE FIRST LOOP

| Process | Algorithm | $K_{p1}$ | $K_{i1}$ | $K_{d1}$ |
|---|---|---|---|---|
| Wood and Berry | GA | 0.252294 | 0.001602 | 0.252587 |
| | ES | 0.25363 | 0.001868 | 0.280028 |
| | CA | 0.273853 | 0.001801 | 0.25349 |
| Vinante and Luyben | GA | -1.20453 | -0.15561 | -0.33433 |
| | ES | -1.26164 | -0.16466 | -0.41223 |
| | CA | -1.21045 | -0.15963 | -0.41414 |
| Wardle and Wood | GA | 4.874161 | 0.047187 | 0.009791 |
| | ES | 5.067988 | 0.045413 | 0.009664 |
| | CA | 4.917404 | 0.045597 | 0.010642 |
| Ogunnaike and Ray | GA | 0.642728 | 0.287408 | -0.02641 |
| | ES | 0.584019 | 0.210688 | -0.03858 |
| | CA | 0.65208 | 0.241827 | -0.01971 |

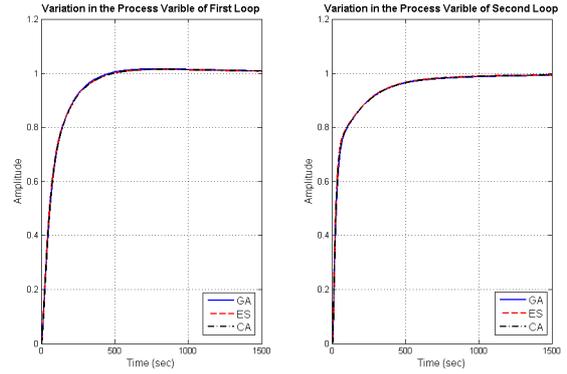

Figure 6. Variation in process variables for WW MIMO process.

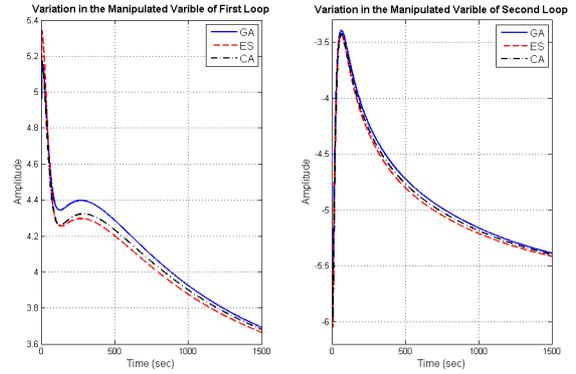

Figure 7. Variation in manipulated variables for WW MIMO process.

TABLE II. TUNED CONTROLLER PARAMETERS FOR THE SECOND LOOP

| Process | Algorithm | $K_{p2}$ | $K_{i2}$ | $K_{d2}$ |
|---|---|---|---|---|
| Wood and Berry | GA | -0.25612 | -0.01075 | -0.52361 |
| | ES | -0.26353 | -0.01248 | -0.53572 |
| | CA | -0.25855 | -0.01177 | -0.5521 |
| Vinante and Luyben | GA | 1.886448 | 0.154075 | 0.713463 |
| | ES | 1.660539 | 0.155043 | 0.544667 |
| | CA | 1.648765 | 0.16293 | 0.572449 |
| Wardle and Wood | GA | -5.37666 | -0.05618 | 0.003258 |
| | ES | -5.58605 | -0.05802 | 0.002066 |
| | CA | -5.54188 | -0.05698 | 0.002082 |
| Ogunnaike and Ray | GA | 0.249743 | 0.148489 | 0.043295 |
| | ES | 0.223773 | 0.142789 | 0.036895 |
| | CA | 0.25331 | 0.144433 | 0.053634 |

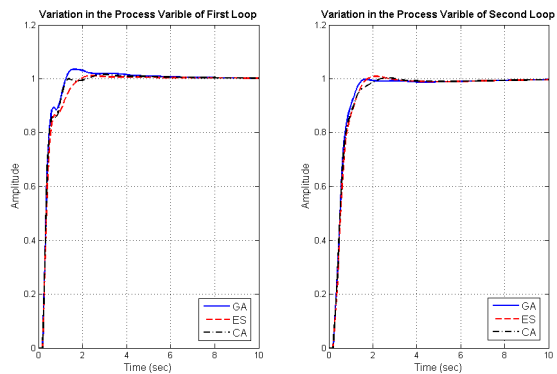

Figure 8. Variation in process variables for OR MIMO process.

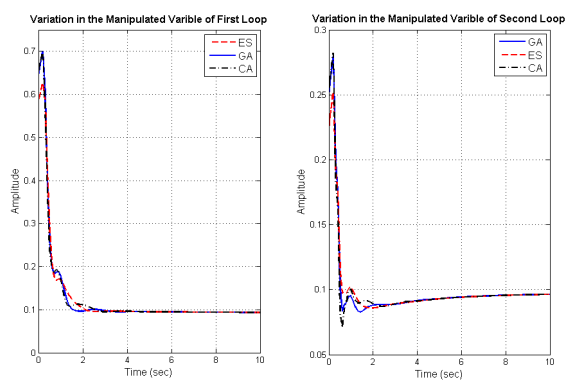

Figure 9. Variation in process variables for OR MIMO process.

TABLE III. CONSISTENCY OF THE EAS FOR MULTI-LOOP PID CONTROLLER DESIGN

| Process | Algorithm | $J_{min}$ (mean) | $J_{min}$ (standard deviation) |
|---|---|---|---|
| Wood and Berry | GA | 14.99845738 | 0.081435328 |
| | ES | 14.88321436 | 0.100647197 |
| | CA | 14.72939843 | 0.031005916 |
| Vinante and Luyben | GA | 34.97645399 | 0.017734663 |
| | ES | 34.93504378 | 0.02309397 |
| | CA | 34.90486417 | 0.003799923 |
| Wardle and Wood | GA | 22386.76045 | 27.45458538 |
| | ES | 22369.29554 | 3.637316428 |
| | CA | 22363.25092 | 0.237556354 |
| Ogunnaike and Ray | GA | 2.18127584 | 0.001161726 |
| | ES | 2.180784464 | 0.00223164 |
| | CA | 2.178030194 | 0.001346768 |

## V. CONCLUSION

In this paper, a new strategy was proposed for the decentralized PID controller tuning for TITO processes. The optimization based controller parameter selection uses an objective function as a weighted sum of ITSE and ISCO. The minimization problem was solved using three popular global optimization techniques viz. Genetic Algorithm, Evolutionary Strategies and Cultural Algorithm for each of the four test-bench TITO processes. It is observed that for each cases, though all algorithms lead to stable controllers, CA was the most effective amongst the three EAs. This is reflected by the standard deviation and mean of the cost functions, confirming the effectiveness of CA. Simple GA is observed to be the most ineffective among the three EAs.